\documentclass[sort&compress,final]{aipproc}\layoutstyle{8x11double}
\usepackage{amsmath}\usepackage{amssymb}\usepackage{booktabs}
\begin{document}\title{Heavy-Quark Mass and Heavy-Meson Decay
Constants from QCD Sum Rules}
\author{Wolfgang LUCHA}{address={Institute for High Energy Physics,
Austrian Academy of Sciences, Nikolsdorfergasse 18, A-1050 Vienna,
Austria}}\author{Dmitri MELIKHOV}{address={Institute for High
Energy Physics, Austrian Academy of Sciences, Nikolsdorfergasse
18, A-1050 Vienna, Austria},altaddress={Faculty of Physics,
University of Vienna, Boltzmanngasse 5, A-1090 Vienna, Austria}}
\author{Silvano SIMULA}{address={INFN, Sezione di Roma III, Via
della Vasca Navale 84, I-00146 Roma, Italy}}

\begin{abstract}We present a sum-rule extraction of decay constants
of heavy mesons from the two-point correlator of heavy-light
pseudoscalar currents. Our primary concern is to control the
uncertainties of the decay constants, induced by both input QCD
parameters and limited accuracy of the sum-rule method. Gaining
this control is possible by applying our novel procedure for the
extraction of hadron observables utilizing
Borel-parameter-depending dual thresholds. For the charmed mesons,
we obtain
$f_D=(206.2\pm7.3_{\rm(OPE)}\pm5.1_{\rm(syst)})\;\mbox{MeV}$ and
$f_{D_s}=(245.3\pm15.7_{\rm(OPE)}\pm4.5_{\rm(syst)})\;\mbox{MeV}.$
In the case of the beauty mesons, the decay constants prove to be
extremely sensitive to the exact value of the $b$-quark
$\overline{\rm MS}$ mass $\overline{m}_b(\overline{m}_b)$. By
matching our sum-rule prediction for $f_B$ to the lattice
outcomes, the very accurate $b$-mass value
$\overline{m}_b(\overline{m}_b)=(4.245\pm0.025)\;\mbox{GeV}$ is
found,~which~yields
$f_B=(193.4\pm12.3_{\rm(OPE)}\pm4.3_{\rm(syst)})\;\mbox{MeV}$ and
$f_{B_s}=(232.5\pm18.6_{\rm(OPE)}\pm2.4_{\rm(syst)})\;\mbox{MeV}.$
\end{abstract}\maketitle

\section{Quark--Hadron Duality}The calculation of the decay
constants $f_P$ of ground-state heavy pseudoscalar mesons $P$ by
QCD sum rules \cite{lms2010,svz} is a complicated problem: First,
a reliable operator product expansion (OPE) for the ``Borelized''
correlation function of two pseudoscalar heavy-light currents has
to be found. Second, even if all the parameters of this OPE are
known precisely, the knowledge of only a {\em truncated\/} OPE for
the correlator allows to extract bound-state observables with only
finite accuracy, reflecting an inherent uncertainty of the QCD
sum-rule approach. Controlling this uncertainty constitutes a
delicate problem for actual applications \cite{lms_1}.

Recall one essential feature of the sum-rule extractions of decay
constants: the quark--hadron duality assumption entails a (merely
approximate) relation between hadronic ground-state contribution
and OPE with the ``QCD-level'' correlator cut at some effective
continuum threshold $s_{\rm eff}$:\begin{align}\label{SR_QCD}
&f_Q^2\,M_Q^4\exp(-M_Q^2\,\tau)=\Pi_{\rm dual}(\tau,s_{\rm eff})\\
&\equiv\hspace{-2.4ex}\int\limits^{s_{\rm
eff}}_{(m_Q+m)^2}\hspace{-2.3ex}{\rm d}s\exp(-s\,\tau)\,\rho_{\rm
pert}(s)+\Pi_{\rm power}(\tau).\nonumber\end{align}Here, the
perturbative spectral density $\rho_{\rm pert}(s)$ is obtained as
a series expansion in powers of the strong coupling $\alpha_{\rm
s}$:$$\rho_{\rm pert}(s)=\rho^{(0)}(s)+\frac{\alpha_{\rm
s}}{\pi\,}\,\rho^{(1)}(s)+\frac{\alpha^2_{\rm
s}}{\pi^2}\,\rho^{(2)}(s)+\cdots.$$Obviously, in order to extract
a decay constant $f_Q$ one has to find a way to fix the effective
continuum threshold~$s_{\rm eff}$.

A crucial albeit very trivial observation is that $s_{\rm eff}$
must be a function of $\tau$, otherwise the l.h.s.\ and the
r.h.s.\ of (\ref{SR_QCD}) would exhibit a different
$\tau$-behaviour. The {\em exact effective continuum threshold\/}
--- corresponding to the true values of hadron mass and decay
constant on the l.h.s.\ of (\ref{SR_QCD}) --- is, of course, not
known. Therefore, our idea of extracting hadron parameters from
sum rules consists in attempting (i) to find a reliable
approximation to the exact threshold $s_{\rm eff}$ and (ii) to
control the accuracy of this approximation. In a recent series of
publications \cite{lms_new}, we have constructed all the
associated procedures, techniques and algorithms.

We define a dual invariant mass $M_{\rm dual}$ and a dual decay
constant $f_{\rm dual}$ ($M_Q$ still denoting the true hadron
mass) by\begin{align}\label{mdual}M_{\rm
dual}^2(\tau)&\equiv-\frac{{\rm d}}{{\rm d}\tau}\log\Pi_{\rm
dual}(\tau, s_{\rm eff}(\tau)),\\ \label{fdual}f_{\rm
dual}^2(\tau)&\equiv M_Q^{-4}\exp(M_Q^2\tau)\,\Pi_{\rm dual}(\tau,
s_{\rm eff}(\tau)).\end{align}In case the ground-state mass $M_Q$
is known, the deviation of the dual ground-state mass $M_{\rm
dual}$ from its actual value $M_Q$ yields an indication of the
excited-state contributions picked up by our dual correlator.
Assuming some specific functional shape for our effective
threshold and requiring least deviation of the dual mass
(\ref{mdual}) from its actual value in the Borel window leads to a
variational solution for the effective threshold. With $s_{\rm
eff}(\tau)$ at our disposal we get the decay constant from
(\ref{fdual}). The standard {\em assumption\/} for the effective
threshold is that it is a ($\tau$-independent) constant. In
addition to such crude approximation we also consider polynomials
in $\tau$. In fact, $\tau$-dependent thresholds greatly facilitate
reproducing the actual mass value. This implies that a dual
correlator with $\tau$-dependent threshold isolates the ground
state much better and is less contaminated by excited states than
a dual correlator with the conventional $\tau$-independent
threshold. As consequence, the accuracies of extracted hadron
observables are drastically improved. Recent experience from
potential models reveals that the band of values obtained from
linear, quadratic, and cubic Ans\"atze for the effective threshold
encompasses the true value of the decay constant \cite{lms_new}.
Moreover, we could show that the extraction procedures in quantum
mechanics and in QCD are even quantitatively very similar
\cite{lms_qcdvsqm}. Here, we report our results
\cite{lms2010,lms2010_conf} for heavy-meson decay constants.

\section{OPE and heavy-quark masses}
\begin{figure}[!b]\begin{tabular}{c}
\includegraphics[width=7.55cm]{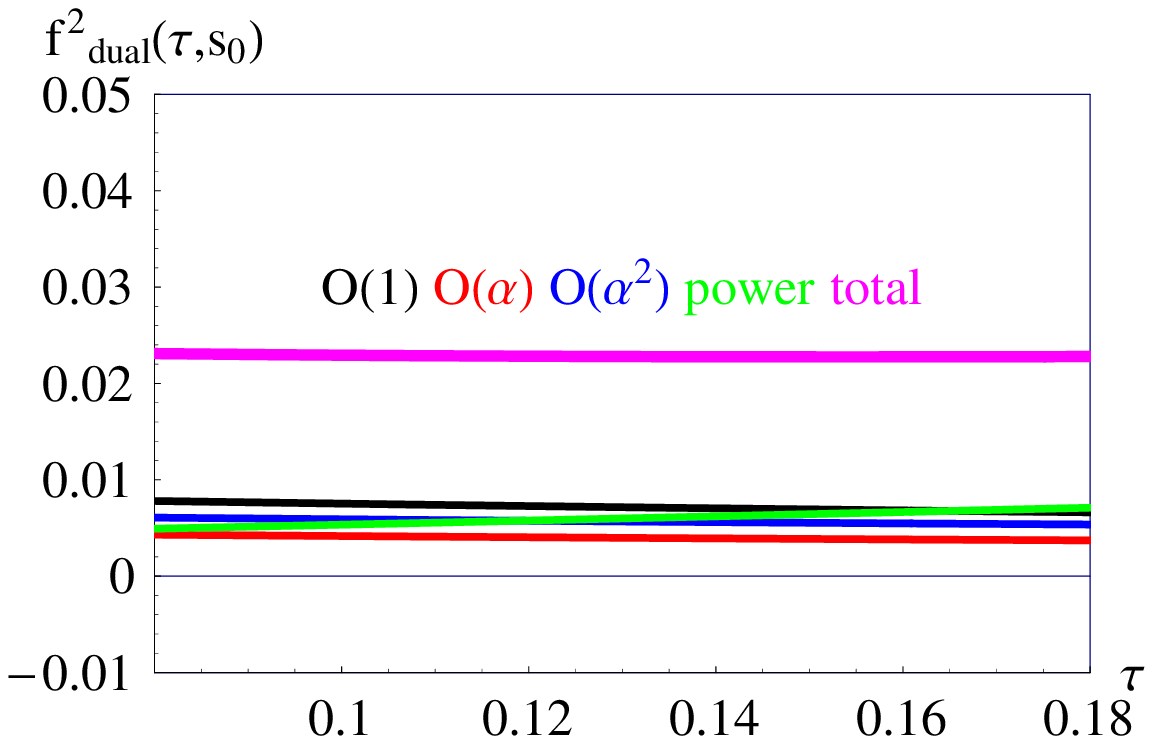}\\
\includegraphics[width=7.55cm]{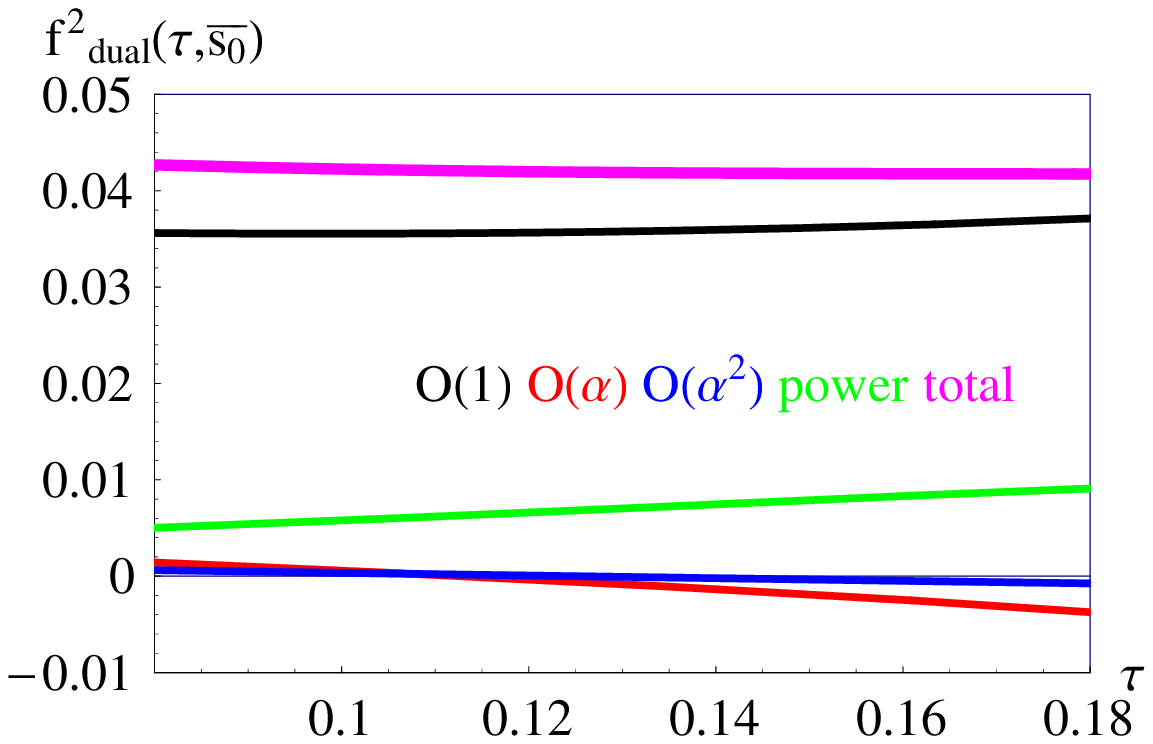}
\end{tabular}\caption{\label{Plot:1}Decay constants $f_B$ extracted
from the correlator given in terms of the $b$-quark pole (top) and
$\overline{\rm MS}$ (bottom) mass.}\end{figure}For heavy-light
correlators and emerging decay constants the choice of the precise
scheme adopted for defining the heavy-quark mass has a great
impact. We utilize the OPE for this correlator to three-loop
accuracy \cite{chetyrkin}, obtained in terms of the pole mass of
the heavy quark. The pole-mass scheme is standard and has been
used for a long time \cite{aliev}. An alternative is to reorganize
the perturbative expansion in terms of the running $\overline{\rm
MS}$ mass \cite{jamin}. Since the correlator is known up to
O$(\alpha^2_{\rm s})$, also the relation between pole and
$\overline{\rm MS}$ mass is applied to such accuracy. Figure
\ref{Plot:1} depicts the resulting $B$-meson decay constant $f_B$
for these two cases. In each case, a {\em constant\/} effective
continuum threshold is fixed by requiring maximum stability of the
found decay constant. Thus, the constant thresholds differ for
pole ($s_0$) and $\overline{\rm MS}$ ($\overline{s_0}$) mass
schemes. Several lessons can be learnt:

\vspace{1ex}(a) In the pole-mass scheme, the perturbative series
for the decay constant shows no sign of convergence: each of the
LO, NLO, NNLO terms contributes with similar size. Consequently,
the pole-mass-scheme result for the decay constant may
significantly underestimate the exact value.

\vspace{1ex}(b) Reorganizing the perturbative series in terms of
the $\overline{\rm MS}$ mass of the heavy quark yields a distinct
hierarchy of the perturbative contributions \cite{jamin}.
Moreover, the absolute value of the decay constant extracted in
this scheme turns out to be some 40\% larger than in the pole-mass
case (a).

\vspace{1ex}(c) Note that, in {\em both\/} cases, the decay
constant exhibits perfect stability in a wide range of the Borel
parameter $\tau$. Thus, mere ``Borel stability'' is {\em not\/}
sufficient to guarantee the reliability of some sum-rule
extraction of bound-state parameters. We have pointed out this
observation already several times \cite{lms_1}. Nevertheless, some
authors still regard Borel stability as a proof of the reliability
of their results.

\vspace{1ex}In the light of our above findings, we adopt in the
next sections the OPE formulated in terms of the $\overline{\rm
MS}$ mass~\cite{jamin}.

\section{Decay constants of $D$ and $D_s$}The application of our
extraction procedures leads to the following values of the
charmed-meson decay constants:\begin{align}
f_{D}&=(206.2\pm7.3_{\rm(OPE)}\pm5.1_{\rm(syst)})\;\mbox{MeV},\label{Dres}\\
f_{D_s}&=(245.3\pm15.7_{\rm(OPE)}\pm4.5_{\rm(syst)})\;\mbox{MeV}.\label{Dsres}
\end{align}

\begin{figure}[!h]\begin{tabular}{c}
\includegraphics[width=7.1cm]{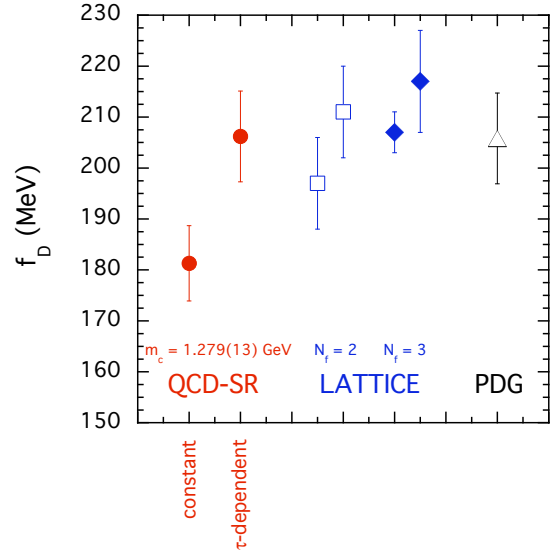}\end{tabular}
\caption{\label{Plot:Dresults}Comparison of our results for the
decay constant $f_D$ with lattice findings. For a detailed list of
references, cf.~\cite{lms2010}.}\end{figure}

\noindent The OPE-related errors in (\ref{Dres},\ref{Dsres}) are
obtained by bootstrap studies allowing for the variation of all
QCD parameters, that is, quark masses, $\alpha_{\rm s}$, and
condensates, in the relevant ranges. We observe perfect agreement
of our predictions with the corresponding lattice results
(Fig.~\ref{Plot:Dresults}). It has to be emphasized that our
$\tau$-dependent effective threshold is a crucial ingredient for a
successful extraction of the decay constant from the sum rule
(\ref{SR_QCD}). Obviously, the (standard) $\tau$-independent
approximation entails a much lower value for the $D$-meson decay
constant $f_D$ that resides rather far from both the experimental
data {\em and\/} the lattice outcome.

\section{Decay constants of $B$ and $B_s$}
\begin{figure}[!h]\begin{tabular}{c}
\includegraphics[width=7.1cm]{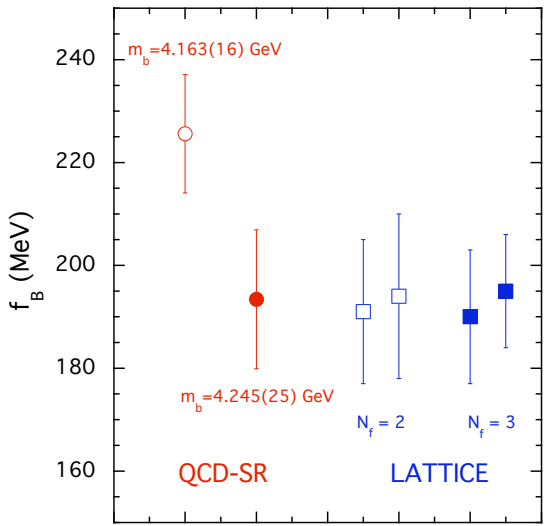}\end{tabular}
\caption{\label{Plot:Bresults}Comparison of our results for the
decay constant $f_B$ with lattice findings. For a detailed list of
references, cf.~\cite{lms2010}.}\end{figure}

Our QCD sum-rule findings
for the beauty-meson decay constants turn out to be extremely
sensitive to the chosen value of the $b$-quark $\overline{\rm MS}$
mass $\overline{m}_b(\overline{m}_b)$. For instance, the range
$\overline{m}_b(\overline{m}_b)=(4.163\pm0.016)\;\mbox{GeV}$
\cite{mb} entails results that are barely compatible with recent
lattice calculations of these decay constants
(Fig.~\ref{Plot:Bresults}). Requiring our sum-rule $f_B$ result to
match the average of the lattice computations provides the rather
precise value of the $b$-quark $\overline{\rm MS}$ mass
$$\overline{m}_b(\overline{m}_b)=(4.245\pm0.025)\;{\rm GeV}.$$For
this value of the $b$-quark mass, our sum-rule estimate for the
$B$- and $B_s$-meson decay constants $f_B$ and $f_{B_s}$ reads
\begin{align*}
f_{B}&=(193.4\pm12.3_{\rm(OPE)}\pm4.3_{\rm(syst)})\;{\rm MeV},\\
f_{B_s}&=(232.5\pm18.6_{\rm(OPE)}\pm2.4_{\rm(syst)})\;{\rm MeV}.
\end{align*}

\section{Summary and Conclusions}Applying (in an attempt to improve
the sum-rule method) our above modifications, we realize several
serendipities:

\vspace{1ex}\noindent 1.~~~The $\tau$-dependence of the effective
thresholds emerges naturally when one attempts to render the
duality relation exact: the dependence is evident from
(\ref{SR_QCD}). We emphasize two facts: (a) In principle, such
dependence on $\tau$ is {\em not\/} in conflict with any
properties of quantum field theories. (b) Our analysis of $D$
mesons shows that it indeed improves {\em decisively\/} the
quality of the related sum-rule predictions.

\vspace{1ex}\noindent 2.~~~Our study of {\em charmed mesons\/}
clearly demonstrates that using Borel-parameter-dependent
thresholds leads to lots of essential improvements: (i) The
accuracy of decay constants predicted by sum rules is drastically
improved. (ii) It has become possible to obtain a realistic
systematic error and to diminish it to the level of, say, a few
percent. (iii) Our prescription brings QCD sum-rule findings into
perfect agreement with both lattice QCD and experiment.

\vspace{1ex}\noindent 3.~~~The {\em beauty-meson\/} decay
constants $f_{B_{(s)}}$ are extremely sensitive to the choice of
the $b$-quark mass: Matching our QCD sum-rule $f_B$ outcome to the
corresponding average of lattice computations provides a truly
accurate estimate of $\overline{m}_b(\overline{m}_b)$, in good
agreement with several lattice results but, interestingly, not at
all overlapping with a recent very accurate determination
\cite{mb} (for details, consult Ref.~[1]).

\vspace{2.7ex}\noindent{\bf Acknowledgments.}~~~D.~M.\ is grateful
for support by the Austrian Science Fund (FWF) under Project
No.~P20573.

\bibliographystyle{aipproc}\end{document}